\documentstyle[12pt,aasms4]{article}

\newcommand{\einstein}{{\it Einstein}\ }

\newcommand{\lx}{$L_{\rm{X}}$}
\newcommand{\lb}{$L_{\rm{B}}$}

\newcommand{\afe}{$A_{\rm{Fe}}$}
\newcommand{\al}{$A_{\rm{\alpha}}$}

\newcommand{\brems}{bremsstrahlung}

\begin{document}
\title{NEW MEASUREMENT OF METAL ABUNDANCE IN THE ELLIPTICAL
GALAXY NGC~4636 WITH ASCA}
\author{{\sc Kyoko} {\sc Matsushita}\altaffilmark{1,2}, {\sc Kazuo} {\sc Makishima}\altaffilmark{2,3}, {\sc Etsuko} {\sc Rokutanda}\altaffilmark{1},
\\ {\sc Noriko} Y. {\sc Yamasaki}\altaffilmark{1}
and {\sc Takaya} {\sc Ohashi}\altaffilmark{1}}
\altaffiltext{1}{Department of Physics, Tokyo Metropolitan University,
1-1 Minami-Ohsawa Hachioji, Tokyo 192-03, Japan; matusita@phys.metro-u.ac.jp}
\altaffiltext{2}{Department of Physics, University of Tokyo,
7-3-1 Hongo, Bunkyo-ku, Tokyo 113, Japan}
\altaffiltext{3}{Research Center for the Early Universe (RESCEU),
University of Tokyo,
7-3-1 Hongo, Bunkyo-ku, Tokyo 113, Japan}

\begin{abstract}

High quality  X-ray spectra of NGC~4636 are obtained  with ASCA. 
Theoretical models are found
 unable to reproduce the data in the Fe-L line region.
Spectral data above 1.4 keV indicate 
that  Mg to Si abundance ratio is $\sim 1$  solar.
Assuming that the abundance ratios among $\alpha$-elements
are the same  with the solar ratios,
spectral fit with increased  systematic error in the 0.4--1.6 keV range
 gives abundances of $\alpha$-elements and Fe to be both $\sim$ 1 solar
 by taking one solar of Fe to be 3.24$\times 10^{-5}$ by number.
 These new abundance results solve  discrepancy between stellar and hot-gas
 metallicity, but still a low supernova rate is implied.
 We also detect  strong abundance gradients for both $\alpha$-elements
 and iron in a similar fashion.
The  abundance is $\sim 1$ solar within $4'$, 
and decreases  outerwards down to $0.2\sim 0.3$ solar  at $10'$ from the
galaxy center.
 Dilution due to an extended hot gas is suggested.
\end{abstract}


 \section{Introduction}

The interstellar medium (ISM) of elliptical galaxies is thought to
be considered to be an accumulation of 
stellar mass loss and supernova ejecta.
While standard supernova rates predict the ISM to have a high metallicity
of several times the solar value (e.g. \cite{loe91}; \cite{ren93}),   
the observed ISM metallicity in fact fell below half a solar, 
even lower than the stellar metallicity 
(e.g. \cite{awaki94}; \cite{mushotzky94};
 \cite{matusita94}; \cite{arimoto}; \cite{matumoto97}).
This strong discrepancy calls into question our current understanding
of supernova enrichment and chemical evolution of galaxies.

NGC 4636 is one of the relatively isolated yet most
luminous elliptical galaxies, 
both in optical (\lb =$ 2.9\times 10^{10} L_\odot$ assuming a distance
of 17 Mpc)  and in X-ray
(\lx = 3.8$\times10^{41} \rm{erg~s^{-1}}$; \einstein) band
(e.g. \cite{forman85};  \cite{catalog}).
Using ROSAT, Trinchieri et al. (1994) discovered a very extended 
X-ray emission surrounding this galaxy, out to $\sim18'$.
The first ASCA observation of NGC 4636 in the PV (Performance Verification) 
phase  yielded the very low ISM metallicity (\cite{awaki94}),
together with abundance and temperature gradients (\cite{mushotzky94}).

We have re-observed NGC 4636 with ASCA (\cite{asca})
for an extremely long time; over 200 ks.
This has allowed Matsushita (1997) 
to perform a much deeper study of the extended X-ray component after ROSAT.
In this paper, we utilize the overall ASCA 
 data including this long exposure, to study the spectral properties of
NGC 4636 and look into the abundance problem.

\section{Observations}

NGC 4636 has been observed from ASCA twice. 
Following the PV observation (1993 July 22, with the SIS in 4CCD mode),
the second much longer observation was conducted in the AO-4 phase
from 1995 December 28 through 1996 January 4, with the SIS in the 1CCD mode.
We discarded the data taken under cut-off rigidities 
less than  6 GeV c$^{-1}$, or elevation angle less than $5^\circ$ 
and $20^\circ$ from night and day earth respectively.
This has yielded exposure times of 36 ks (with the GIS) 
and 39 ks (with the SIS) for the PV observation, and
those for the AO-4 observation are 172 ks (GIS) and 215 ks (SIS).

We accumulated on-source spectra within 4 times 
the effective radius, $r_e=1'.7$ (\cite{faber}), centered on NGC~4636. 
Since the SIS response has changed with time,
we  treat the SIS spectra from the two observations separately.
The background spectrum  was obtained by integrating 
the blank-sky data over the same region of the detector.
Figure \ref{fig1} shows the background-subtracted SIS spectrum
for the AO-4 data.

\section{Results}

We jointly fit the two SIS (PV and AO-4) spectra 
and one GIS
 spectrum, with a standard two component model
(\cite{awaki94}; \cite{matusita94}; \cite{matumoto97}).
The model consists of a thin thermal emission from the ISM
with free temperature $kT$ and free metallicity, 
and a thermal \brems~ with temperature fixed at 10 keV 
representing the contribution from low-mass X-ray binaries.
Both components are subjected to a common interstellar absorption $N_{\rm H}$.
   Although the GIS  (\cite{ohashigis}; \cite{maxgis}) is less sensitive than the SIS to the low-energy
(e.g. $<$ 1 keV)
   atomic lines, it can constrain the hard component better than the SIS,
   and is fully sensitive to the Si-K and S-K lines. Therefore, the joint use
   of the two instruments is essential.
In this paper we adopt for the solar iron abundance the `meteoritic' value,
 Fe/H$=3.24\times 10^{-5}$ by number (\cite{feh})

\subsection{Problems with the spectral analysis}

As the first-cut spectral analysis, we represented the ISM 
component by the plasma emission model of Raymond \& Smith 
(1977; hereafter R-S model) with solar abundance ratios.
The best-fit model parameters turned out to be consistent with 
those of Awaki et al. (1994), but the fit was totally unacceptable (Table 1).
We then allowed the abundances to deviate from the solar ratios, 
by dividing heavy elements into two groups to estimate relative
contributions from type Ia and type II supernovae (SNe).
One group consists of so-called $\alpha$-elements, O, Ne, Mg, Si and S,
which are assumed to have a common abundance $A_{\alpha}$.
The other group includes Fe and Ni, with a common abundance $A_{\rm{Fe}}$.
Abundance of He is fixed to be 1 solar.
The abundances of the other elements are assumed to be the same as
 $\alpha$-elements, although their effect on $1$ keV spectrum is
negligible.
The fit incorporating 6 parameters (\al, \afe, $kT$, $N_{\rm H}$, 
and two normalizations) still remained far from acceptable, as shown in Table 1.
We further replaced the R-S model with MEKA model (\cite{mewe85}; \cite{mewe86};
\cite{kaastra}), 
MEKAL model (\cite{liedahl95}), or Masai model (\cite{masai84}), 
but none of them were successful.
Multi-temperature ISM models did not improve the fit, either.

In these fits, the data-to-model discrepancy is always largest 
around the Fe-L complex region (0.8 $\sim 1.4$ keV).
In addition, the fit residuals and the derived physical quantities 
both depend significantly on the plasma emission codes. 
These reconfirm the serious problems in the theoretical 
Fe-L line modeling, as pointed out by Fabian et al. (1994)
 and Arimoto et al. (1997).
Furthermore, we have found strong false couplings 
between $A_{\alpha}$ and $A_{\rm{Fe}}$, 
arising from the following two effects.
On one hand, the fitting algorithm tries to reduce the Fe-L discrepancy,
by adjusting intensities of the O-K and Ne-K lines
which overlap the Fe-L complex; this strongly affects $A_{\alpha}$,
since the data have the highest statistics in this energy range. 
In turn $A_{\alpha}$ affects $A_{\rm Fe}$,
because the bound-free emission from oxygen and neon acts 
nearly as a continuum to the Fe-L lines and controls their equivalent widths.
These effects make both \afe~ and \al~ highly unreliable.

\subsection{Abundance of $\alpha$-elements}

In order to avoid these problems,
we tentatively restricted the  energy range for the SIS spectral fit
to $>1.4$ keV which is unaffected by the Fe-L, O-K and Ne-K lines.
For the GIS spectrum, we used an energy range above 1.6 keV, because of
its poorer energy resolution.
We re-arranged heavy elements into 4 groups; Mg; Si; S;
and other elements including O, Ne, Fe and Ni (hereafter $A_{\rm others}$).
Elements in the last group contribute to the $> 1.4$ keV spectrum
mainly via free-bound continua (Fe-K and Ni-K lines are undetectable).
Then, the  fit has become acceptable with the reduced
$\chi^2$ of 1.06 for 220 dof..
The derived temperature, 0.73 keV (0.67--0.76 keV  with 90\% confidence
for a single parameter), agrees with those derived
 using the whole energy range.
We have further found that the Mg to Si abundance ratio 
always agrees with the solar ratio (Fig.2),
even though their abundances correlate considerably with $A_{\rm others}$.
The S abundance may be similar, or slightly higher by about 30 \%.
On the other hand, the stellar and supernova nucleosysnthesis predicts 
O, Ne and Mg to follow the solar abundance ratios (\cite{sn96}).
Therefore, it is reasonable to assume that all the $\alpha$-elements
(O, Ne, Mg, Si and S) have the same abundance in solar units.

We can now safely return to the original element grouping 
into \al~ and \afe, and refit the spectra above 1.6 keV (SIS) and 1.7 keV (GIS).
The fit is acceptable  
with the reduced $\chi^2$ of 1.06 for 210 dof.,
and the derived temperature is 0.75 (0.72--0.78) keV.
Now, \al~ is determined relatively well (Fig.3),
because it relies upon the Si-K and S-K lines
instead of using the unreliable O-K and Ne-K lines that are now ignored.
Unlike the full-range fit, \al~  depends on \afe~ 
only weakly through Fe contribution to the free-bound continuum.
These results unambiguously show
that \al~ is roughly one solar, unless \afe~ is extremely high.

\subsection{Iron abundance}

Our next task is to better constrain $A_{\rm Fe}$.
This requires us to utilize the Fe-L line information, 
but simply reviving the energy range below 1.6 keV would bring us 
back to the original problems described in \S3.1.
In order to relax the too strong constraints 
imposed by the data in the Fe-L region,
we have decided to assign 20\% systematic errors
to each spectral data in the 0.4--1.6 keV range.
This is because the data-to-model discrepancy and the model-to-model 
differences are both $\sim $ 20 \% around the Fe-L region
(\cite{masai97}).
We have thus fitted the SIS/GIS spectra jointly with the two-component model,
incorporating the increased systematic error 
and again employing the element grouping into \al~ and $A_{\rm Fe}$.

The fit is acceptable with either of the plasma models (Table 1),
and the confidence regions on the \al~ vs. \afe~ plane
for different plasma emission codes
have come to a  similar region  (Fig.3).
As a consequence, we can now conclude with confidence
that \al~ and \afe~ are both consistent with $\sim $ 1 solar, 
within a factor of 2.
In addition, the ratio between \al~ to \afe~ coincides
with the solar ratio within $\sim 20$\%.

\subsection{Metal distribution in the ISM}

In order to study the metal distribution, 
we accumulated the GIS and  SIS spectra in three
concentric annular regions with radius range, 0--4$'$,
4$'$--8$'$, and 8$'$--12$'$ centered on the galaxy. 
We assume that the ISM is spherically symmetric.
This ``ring-sorted analysis'' is however limited by the spectral mixing effect 
among different sky regions, due to the extended point-spread-function
 of the ASCA X-ray Telescope (XRT;  \cite{xrt}).
This tends to smear out any radial gradient in the ISM properties.
Following the analysis in \S3.3, we include 20\% systematic error in the
0.4-1.6 keV  data.
We fit only with the R-S model, since the difference due to plasma emission
code is only several tens of \% after inclusion of the systematic error.
An acceptable fit is obtained in all three regions (Table 2).

As shown in Table 2 and Fig.4,  the derived  \al~ and \afe~ are $\sim$ 1 solar
 within $r<4'$, and drop to $0.2\sim0.3$ solar at $r\sim 10'$.
 This  confirms
the existence of strong abundance gradient in the ISM in NGC 4636,
and reveals that both $\alpha$-elements and Fe equally show the
abundance gradient. Their ratio is close to the solar value at all radii.
The absolute values of the abundances are higher than
those derived in Mushotzky et al. (1994), who assumed a solar
ratio in the metals, but 
the  temperatures and hydrogen column densities are consistent.

\section{Discussion}
By carefully examining the abundance ratios and uncertainties
in the Fe-L complex, we have concluded that the ISM abundances of
Fe and $\alpha$-elements in the X-ray
luminous galaxy, NGC 4636, are in fact as high as about 1 solar within $r<4'$.
The abundance ratio between \al~ and \afe~ is very close to the solar
ratio within $\sim 20$\%.
This means that both  Type Ia and Type II SNe ejecta make significant
contribution in the ISM. 

The ISM iron abundance   is expected to be given by
$$Z^{\rm Fe}_{\rm ISM}= <\! Z^{\rm Fe}_*\!> + 
    5\vartheta_{\rm SN~I}\left( \frac{M^{\rm Fe}}{0.7 M_\odot}\right)h^2_{50},
\eqno(1)$$
(\cite{loe91}; \cite{ren93}),
where $<\! Z^{\rm Fe}_*\!>$ is the average iron abundance of the stars
in units of the solar abundance, $h_{50} \equiv H_0/50$ the Hubble
constant in units of 50 km s$^{-1}~\rm{Mpc}^{-1}$,  $\vartheta_{\rm SN~I}$ 
is the present rate of SN Ia in ellipticals in units of the rate as 
estimated by Tammann (1982), and $M^{\rm Fe}$ is the iron yield
per SN Ia event.
The stellar Fe abundance, $<\! Z^{\rm Fe}_*\!>$, averaged over the whole galaxy 
is  0.74 solar (\cite{arimoto}).
Since our revised value of \afe~  is
fairly close to these stellar values,
the previously-quoted severe problem that
 the ISM abundance was lower than the  stellar metallicity
has been mostly removed.

We can solve for the
relative contribution from SN Ia and SN II on $\alpha$-elements and Fe.
If we assume that SN II products have 3 times higher value of \al~
than \afe~ (\cite{sn96}) and SN Ia produces only Fe,
the SN II contribution necessary to produce the observed \al~
(1 solar) can enrich Fe to \afe~ = 1/3 solar. The rest of
\afe, i.e. 2/3 of the best-fit Fe abundance, has to be explained by
SN Ia.
Using equation (1), we can estimate the SN Ia rate $\vartheta_{\rm SNI}$.
In order to derive an upper limit for $\vartheta_{\rm SNI}$, let us
assume that stars contain no Fe from SN Ia. This gives an equation
only for the SN Ia contribution,
$Z^{\rm Fe}_{\rm ISM,SN Ia} >
    5\vartheta_{\rm SN~I}\left(M^{\rm Fe}/{0.7 M_\odot}\right)h^2_{50}$.
The left hand side is now estimated to be $\sim$ 0.7 solar,
so that the condition for $\vartheta_{\rm SNI}$ is
$\vartheta_{\rm SNI}h^2_{50}< 0.14$.
Therefore, the present abundance result suggests that SN Ia rate
would be less than 1/7 of Tammann's rate for $h_{50}=1$.
Considering the uncertainties in  \al ~and \afe,
this is  consistent with the recent estimation of $\vartheta_{\rm SNI} \sim
1/4$ by Capperallo et al. (1993) based on the supernova counting.
A systematic study of the ISM abundance involving many galaxies
would be neccessary to further look into this problem.
Also, Fe-L diagnostics need to be closely examined in order to
reach consistent answers among the models (\cite{arimoto}).

We have confirmed and refined the strong abundance gradient feature in the ISM
discovered by Mushotzky et al. (1994).
In addition, we have for the first time
 discovered strong abundance gradient in
$\alpha$-elements, which is very similar to that of \afe.
There may be an internal metallicity gradient in the ISM
within the galaxy, as suggested from the significant gradient in the
stellar metallicity within $r_e$ (\cite{mg2grad}).
However, no information is available about the stellar metallicity
in larger scales.
There exists an extended X-ray emission around NGC~4636 with a radius
$>$ 100 kpc, as  discovered by Trinchieri et al. (1994),
and  confirmed with ASCA, by Matsushita (1997).
Since the hot gas responsible for the extended emission
has a very low metallicity as seen in the data in r=8--12$'$,
 the abundance gradient can be 
explained in terms of  the growing contribution from the
 metal poor gas in the outer region.
The similarity in the slope of
 of \al~ and \afe~ gradient  suggests common dilution due to external
gas.

\acknowledgements

The authors are grateful to K. Matsuzaki who have made possible
the exceptionally long observation from ASCA with
incredibly efficient command arrangements.

\newpage

\figcaption[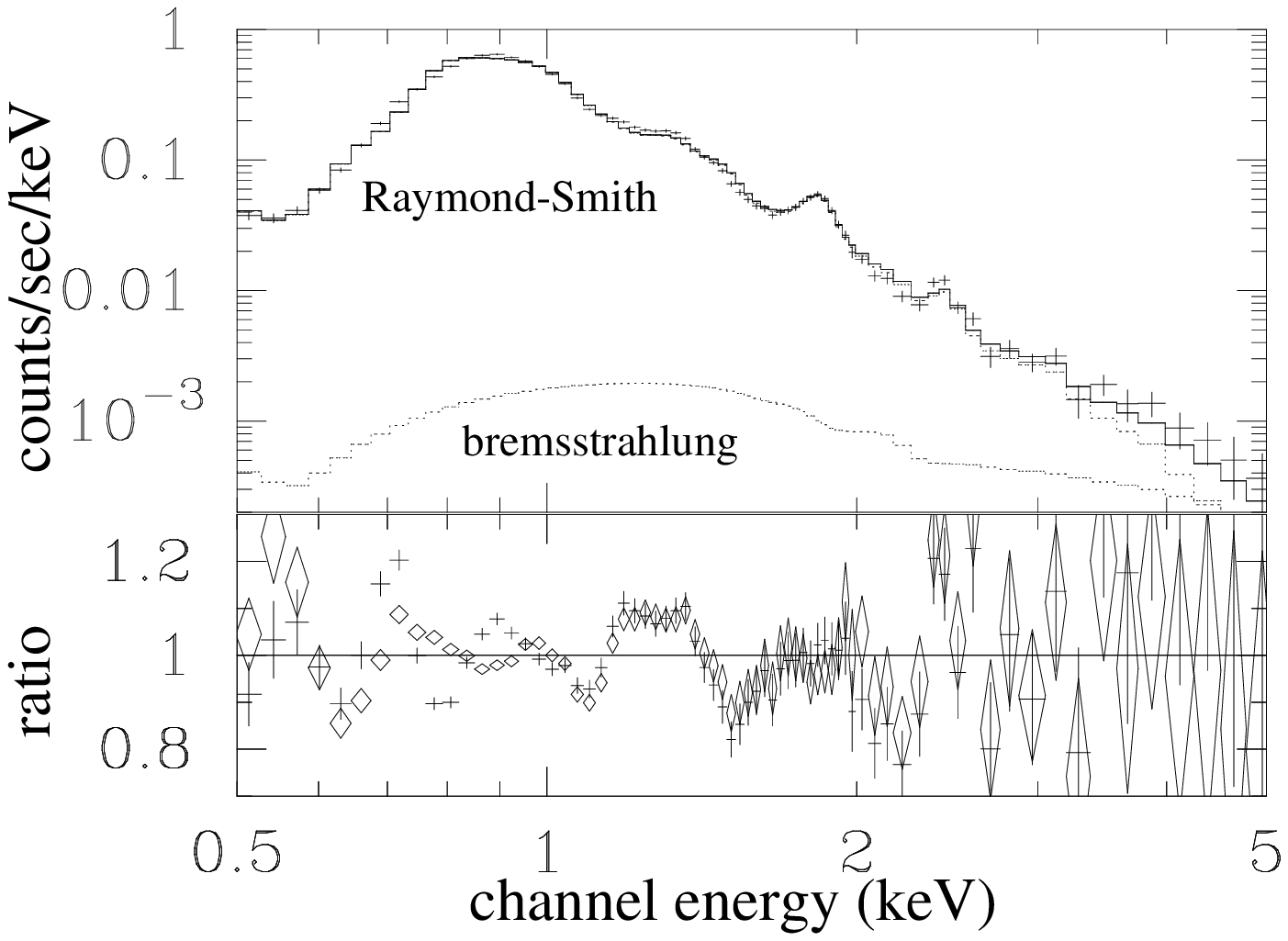]{The SIS spectrum (AO-4 observation) of NGC~4636
 fitted with a two-component model, consisting of a soft
 R-S component (dotted lines) and a hard \brems~ component with $kT$=10 keV
 (dotted line). Abundance of  heavy elements are grouped into $\alpha$-elements
 and iron group.
 The bottom panel shows  data-to-model ratios
 for R-S model (crosses) and MEKAL model (diamonds).
\label{fig1}}

\figcaption[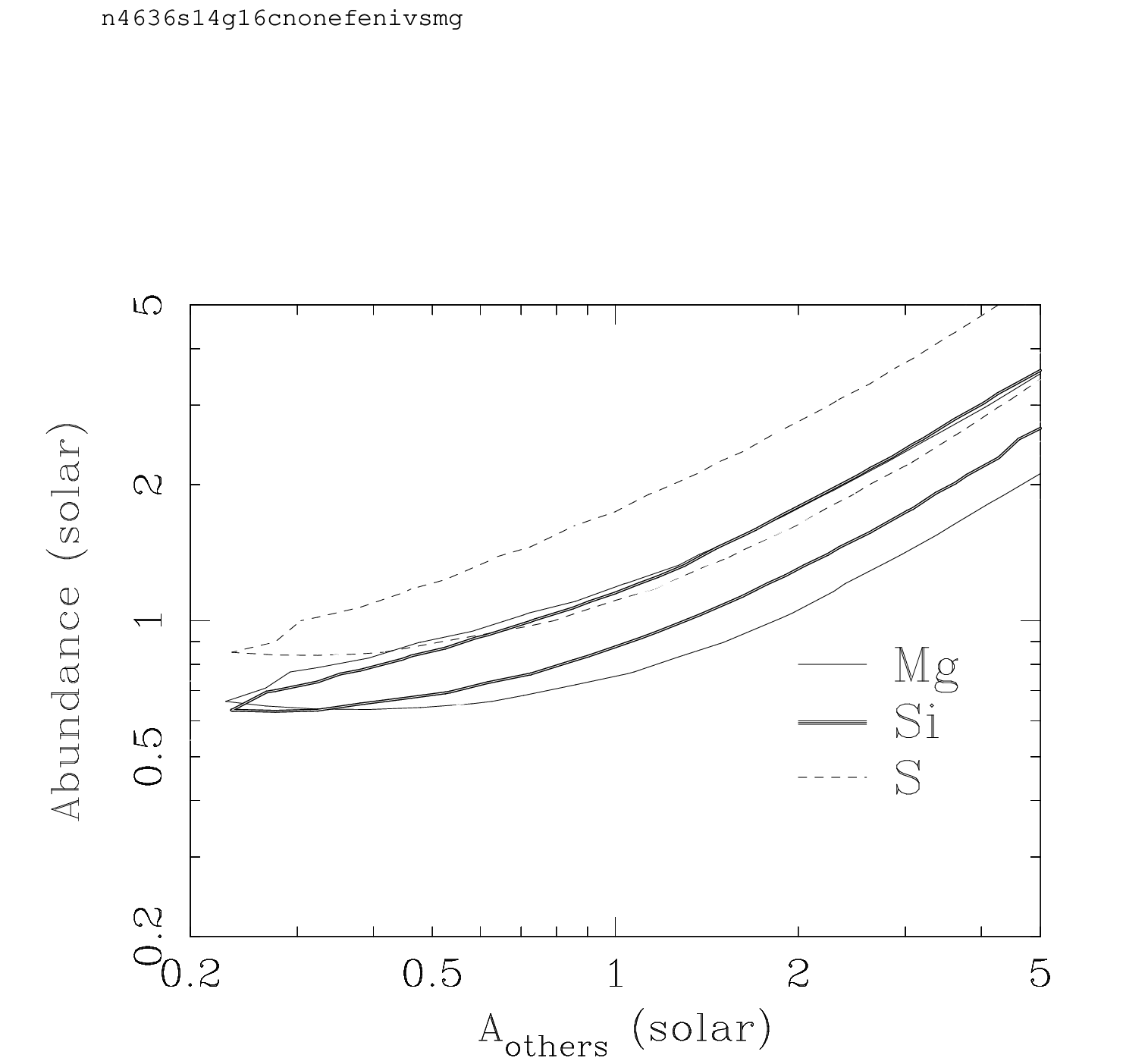]{ Two-parameter
 90\%  confidence contours of abundances of
 Mg (thin line), Si (bold line), and S (dashed lines), shown against the
 $A_{\rm{others}}$.
 These were   obtained from the joint  fit to the spectra above
 1.4 keV (SIS) and 1.6 keV (GIS).
 \label{fig2}}
\figcaption[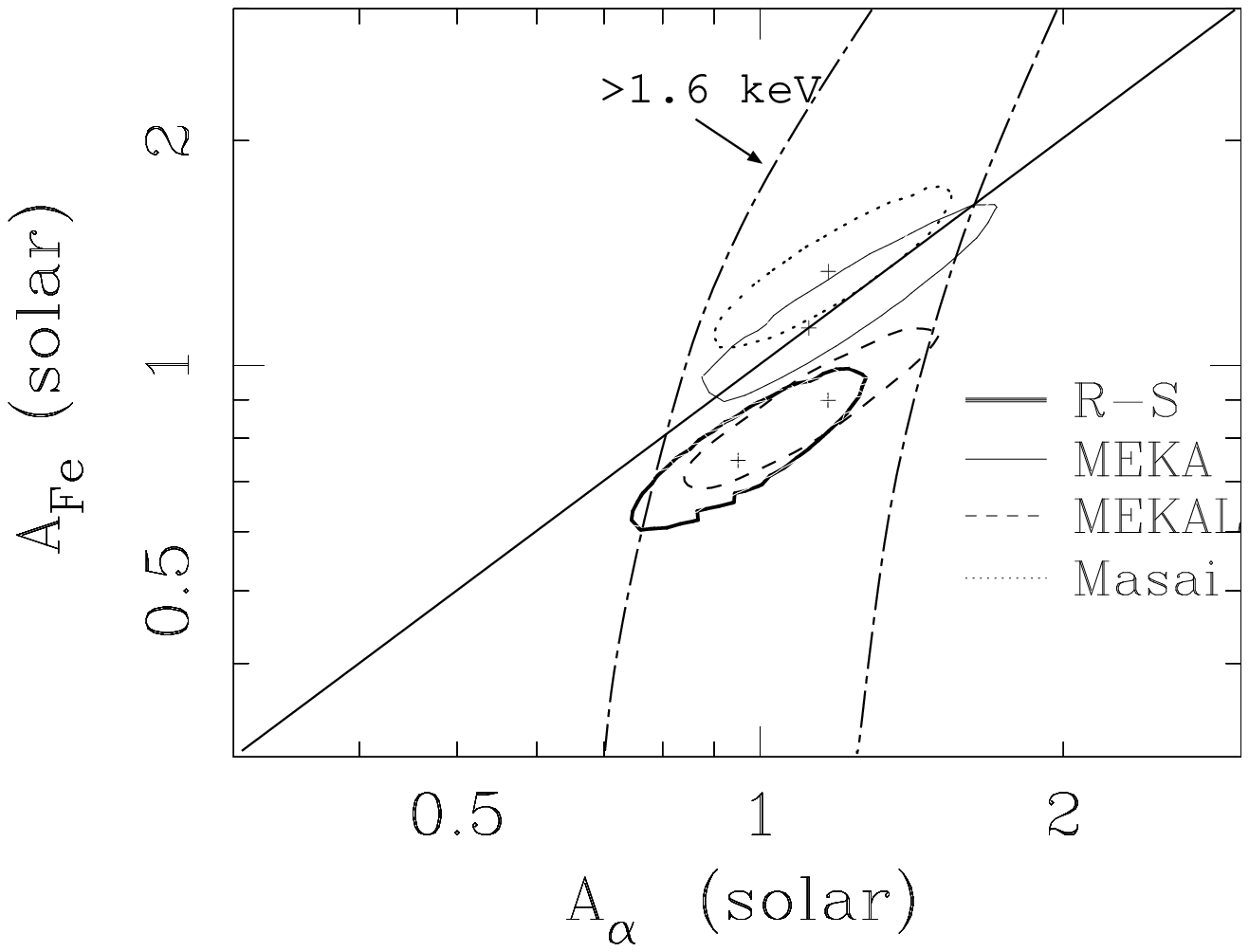]{
 90 \% confidence contour of the correlation
  between \afe~ vs. \al~ obtained by fitting the
 spectra above 1.6 keV (SIS) and 1.7 keV (GIS), using the R-S model (dot-dash),
and those  obtained by R-S (bold), MEKAL (dash),
 MEKA (thin) and Masai (dot) models fitting the whole energy band,
 including  20 \% systematic errors  in the energy range  0.4--1.6 keV.
\label{fig3}}

\figcaption[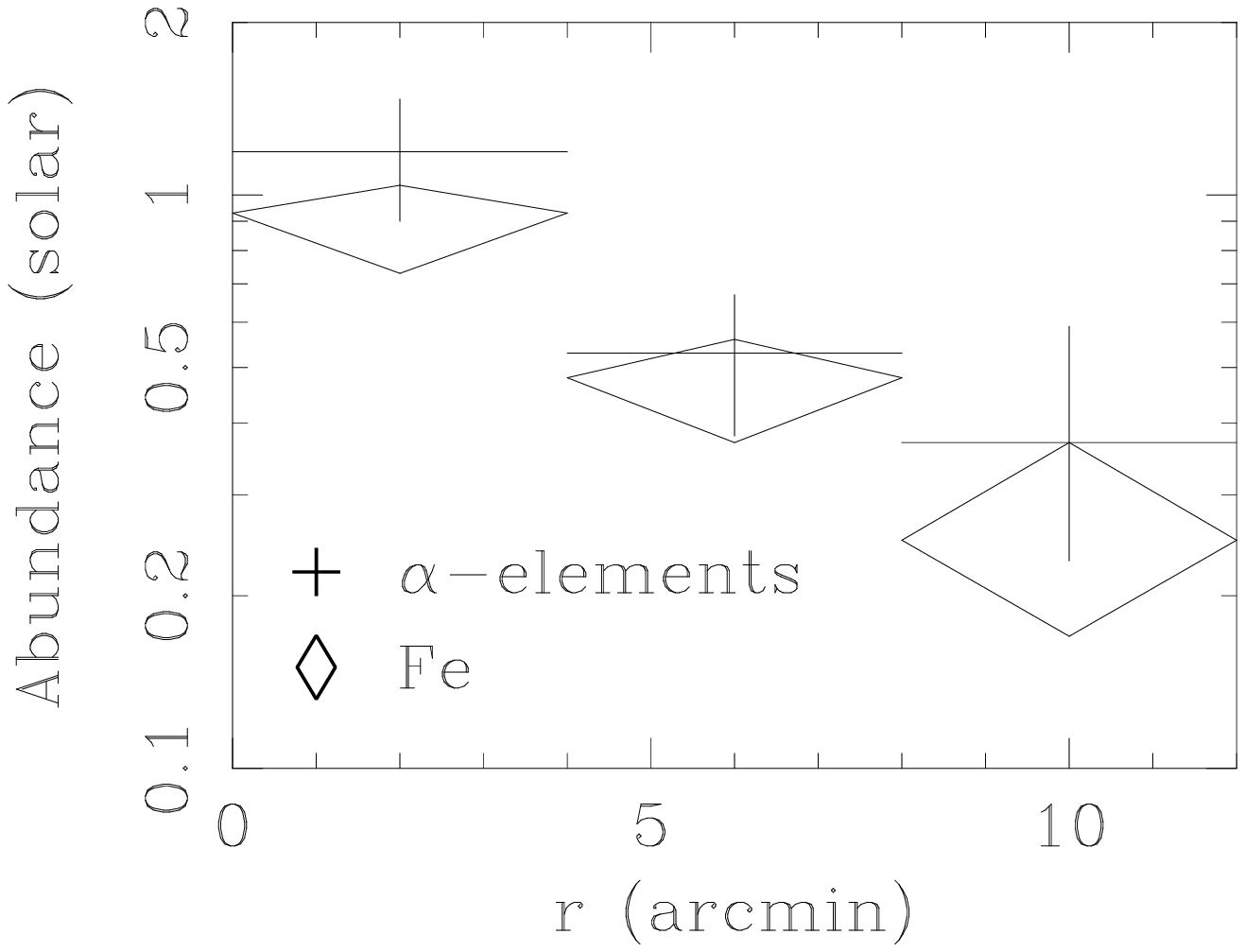]{
Projected radial  profiles of  \al~ (crosses) and \afe~ (diamonds).
\label{fig4}}

\newpage

\begin{table*}
 \begin{center}
\begin{tabular}{lllllr}
  Code & $kT$    & \al & \afe & $N_H$ & $\chi^2/\nu$\\
  &(keV)& (solar)  & (solar) & ($10^{20}\rm{cm^{-2}}$) & \\
 \tableline
 R-S & 0.76 & 0.48     & $=$\al & 5& 1112/373 \\
 \tableline
 R-S & 0.75 & 0.54 & 0.50 &4 & 1093/372\\
MEKA & 0.59 & 1.66 & 1.49  &9 & 1235/372\\
 MEKAL & 0.66 & 0.76 & 0.71 &10 & 727.5/372\\
 Masai & 0.74 & 1.50 &  1.68  & 1 & 786.6/372\\
 \tableline
\multicolumn{6}{c}{Including systematic error in Fe-L region (0.4--1.6 keV) of
 the SIS spectra}\\
\tableline
 R-S & 0.76 $^{+ 0.02 }_{- 0.02 }$ & 0.95 $^{+ 0.25 }_{- 0.15 }$
 & 0.75 $^{+ 0.13 }_{- 0.09 }$ &1 $^{+ 3 }_{-1 }$ & 223.0/330\\
 MEKA & 0.63 $^{+ 0.02 }_{- 0.02 }$ & 1.10 $^{+ 0.44 }_{- 0.21 }$
 & 1.12 $^{+ 0.37 }_{- 0.20 }$ &6 $^{+ 1 }_{- 2 }$ & 337.4/330\\
 MEKAL & 0.72 $^{+ 0.01 }_{- 0.02 }$ & 1.13 $^{+ 0.16 }_{- 0.19 }$
 & 0.88 $^{+ 0.10 }_{- 0.13 }$ &4 $^{+ 3 }_{- 3 }$ & 247.4/330\\
 Masai & 0.72 $^{+ 0.02 }_{- 0.02 }$ & 1.17 $^{+ 0.25 }_{- 0.20 }$
 & 1.33 $^{+ 0.30 }_{- 0.23 }$ &2 $^{+ 2 }_{- 2 }$ & 271.5/330\\
 \tableline
\end{tabular}
 \end{center}
\caption{Summary of joint GIS and SIS spectral fits for the NGC~4636 data
  accumulated within $r<4 r_e$ (6$'$.8). Errors show 90\% confidence
 limits for a single parameter.
 \label{tbl1}}
 \tablenotetext{a}{The definition of solar iron abundance:
 Fe/H=3.24$\times10^{-5}$ (by number)}
\tablenum{1}

\end{table*}

\begin{table*}
\begin{center}
\begin{tabular}{cllllr}
  r & $kT$    & \al & \afe & $N_H$ & $\chi^2/\nu$\\
 (arcmin) &(keV)& (solar)  & (solar) & ($10^{20}\rm{cm^{-2}}$) & \\
 \tableline
 0--4 & 0.75 $^{+0.02}_{-0.02}$&1.19 $^{+0.28}_{-0.29}$  &0.93 $^{+0.11}_{-0.20}$ & 2 $^{+3}_{-2}$& 182.5/264\\
  4--8 &0.81 $^{+0.02}_{-0.03}$&0.53 $^{+0.14}_{-0.15}$ &0.48 $^{+0.08}_{-0.11}$ & 1 $^{+4}_{-1}$ & 202.3/259 \\
  8-12 &0.77 $^{+0.05}_{-0.04}$&0.37 $^{+0.22}_{-0.14}$ &0.25 $^{+0.12}_{-0.08}$  & 0 $^{+2}_{-0}$ & 149.0/194 \\
 \tableline
\end{tabular}
\end{center}
\caption{The fitting results for the ring-cut SIS spectra by adding 
 20 \% systematic errors in   0.4-1.6 keV.
 The Raymond-Smith model has been used. \label{tbl2}}
\tablenum{2}

\end{table*}

\end{document}